\documentclass[aps,prl,twocolumn,showpacs,superscriptaddress,groupedaddress]{revtex4}  
\usepackage{graphicx}  
\usepackage{dcolumn}   
\usepackage{bm}        
\usepackage{amssymb}   

\usepackage{hyperref}
\usepackage[usenames,dvipsnames]{color}
\usepackage{graphicx}
\usepackage{amsmath}
\usepackage{amsfonts}
\usepackage{dsfont}
\usepackage{dcolumn}
\usepackage{bm}
\usepackage{slashed}
\usepackage{pstricks}
\usepackage{slashed}
\usepackage{multirow}
\usepackage{array}
\usepackage{rotating}
\usepackage{xcolor}
\usepackage{epstopdf}
\usepackage{fancybox}
\usepackage{fancyvrb}
\usepackage[normalem]{ulem}
\usepackage{color}
\usepackage{mathdots}
\usepackage{mathrsfs}
\usepackage{multirow}
\usepackage{esint}
\allowdisplaybreaks

\DeclareGraphicsExtensions{.jpg,.pdf,.mps,.png,}

\newcolumntype{x}[1]{
{\centering}p{#1}}%

\newcommand{\GeV}      {~\mathrm{GeV}}

\newcommand{\beqn}{\begin{eqnarray}}
\newcommand{\eeqn}{\end{eqnarray}}
\newcommand{\be}{\begin{equation}}
\newcommand{\ee}{\end{equation}}

\newcommand{\mathsym}[1]{{}}

\newcommand{\st}{Stueckelberg~}

\def \n34{\tilde{\chi}^{0}_{3,4}}

\def\met100{\slashed{E}_T\geq 100 \GeV}

\def\met{\slashed{E}_{T}}

\begin{document}

\title{Baryogenesis from dark matter}

\author{Wan-Zhe~Feng}
\affiliation{
Department of Physics and Institute for Advanced Study, The Hong Kong University of Science and Technology,
Kowloon, Hong Kong SAR, China}
\author{Anupam~Mazumdar}
\affiliation{Consortium for fundamental physics, Lancaster University, LA1 4YB, United Kingdom}
\author{Pran~Nath}
\affiliation{Department of Physics, Northeastern University, Boston, MA 02115, USA}


\begin{abstract}
We consider the possibility that some primordial fields decay
purely into the dark sector creating asymmetric dark matter.
This asymmetry is subsequently
transmuted into leptons and baryons.
Within this paradigm we compute the
amount of asymmetric dark matter created from  the out of equilibrium decays of the primordial fields
with CP violating Yukawa couplings.
The dark matter asymmetry is then transferred to the  visible sector
by the asymmetry transfer equation and generates an excess of $B-L$.
Baryogenesis occurs via sphaleron processes which conserve $B-L$ but violate $B+L$.
A mechanism for the annihilation of the symmetric component of dark matter is also discussed.
The model leads to multi-component dark matter
consisting of both bosonic and fermionic components.\\

\end{abstract}

\maketitle
{\it Introduction:\label{a}}
In this work we discuss the possibility that  a primordial field decays
into the dark sector and creates an asymmetry,
instead of decaying into Standard Model particles.
The dark matter asymmetry  then transmutes into leptons and baryons.
As is well known the generation of a  baryon excess (or a lepton excess) requires satisfaction of the
three Sakharov conditions~\cite{Sakharov:1967dj}: a violation of baryon (or lepton) number, existence of C and CP violation,
and non-equilibrium processes.  Baryon and lepton number violations appear in the Standard Model
and its supersymmetric extensions via higher dimensional operators (for a review see~\cite{Nath:2006ut})
and they  can also arise via spontaneous breaking~\cite{Dulaney:2010dj}. Thus in the analysis of the genesis
of dark matter via the decay of the primordial fields, we will assume the existence of such violations.
The remaining Sakharov conditions are also met leading to the generation of asymmetric dark matter.
A part of dark matter then transmutes to the visible sector. An analysis in similar spirit
where dark matter is the genesis of visible matter was discussed in~\cite{Buckley:2010ui}.
Our model is significantly different from this work and further
we also discuss the genesis of asymmetric  dark matter as arising from the decay of the primordial fields.
The symmetric component of dark matter is depleted by mechanisms similar to those discussed
in~\cite{Feng:2012jn}.
\\

{\it The model:\label{b}}
We will work in a supersymmetric framework where the  superpotential of the model is given by
\beqn
W=
W_{\rm gen} + W_{\rm tran} + W_{\rm MSSM}\,.
\label{WL}
\eeqn
Here
$W_{\rm gen}$ generates the asymmetry for the dark
matter through  the decay of the primordial fields,
$W_{\rm tran}$ transfers the asymmetry from the dark sector to the visible sector, and
$W_{\rm MSSM}$ is the superpotential of the minimal supersymmetric standard model.
Without going into details we note in passing that a possible candidate for the primordial field is an inflaton
(for reviews see \cite {Enqvist:2003gh,Mazumdar:2010sa}).
\\

{\it Genesis of asymmetric dark matter:}
We assume there exist several $\hat{N}_i$ fields ($i \geq 2$) in the early universe with masses $M_i$,
where $\hat N=(N, \tilde{N})$ and $N$ is the Majorana field and $\tilde N$ is the super-partner field.
The scalar field of the lightest $\hat{N}_i$ superfields
could play the role of the inflaton field, i.e., $\phi \equiv \tilde{N}_1$.
The dynamics is driven by the coupling of
the superfields $\hat{N}_i$ to the dark sector.
The dark sector is comprised of $(\hat X, \hat{X}^c, \hat X', \hat{X}'^c)$ which are charged under the gauge group $U(1)_x$ with charges $(+1,-1,-1,+1)$.
All of the MSSM fields are not charged under this new gauge symmetry $U(1)_x$.
We assume the $\hat{N}_i$ superfields carry a non-vanishing lepton number $+2$,
 $\hat X, \hat X'$ carry lepton number $-1$ and $\hat{X}^c, \hat{X}'^c$ carry lepton number $+1$.
$W_{\rm gen}$ is invariant under both $U(1)_x$ and lepton number~\footnote{
In the model only the mass term of the $\hat{N}_i$ superfields ($W \sim \frac{1}{2} M_i \hat{N}_i \hat{N}_i$)
violates the lepton number.}
and it takes the following form
\beqn
W_{\rm gen}=
\lambda_i \hat{N}_i \hat{X} \hat{X}'
+ m \hat X \hat{X}^c + m' \hat X' \hat{X}'^c
\,,
\label{Wgen}
\eeqn
where the Yukawa coupling $\lambda_i$ is assumed to be complex.
The interaction of $W_{\rm gen}$ describes the genesis of dark matter.
It gives rise to the decays
\beqn
N_i \to X \tilde X', \tilde X X', \bar{X} \tilde X'^*, \tilde X^* \bar{X}',
\quad \tilde{N}_i \to  X X',   \bar X \bar X'.
\eeqn
In the simplest model we have $i=2$,
and we assume $\hat{N}_2$ mass $M_2$ is much larger than $\hat{N}_1$ mass $M_1$,
so the generation of the asymmetry in the dark sector is mostly through $\hat{N}_1$.\\

\begin{figure}[t!]
\begin{center}
\includegraphics[scale=0.54]{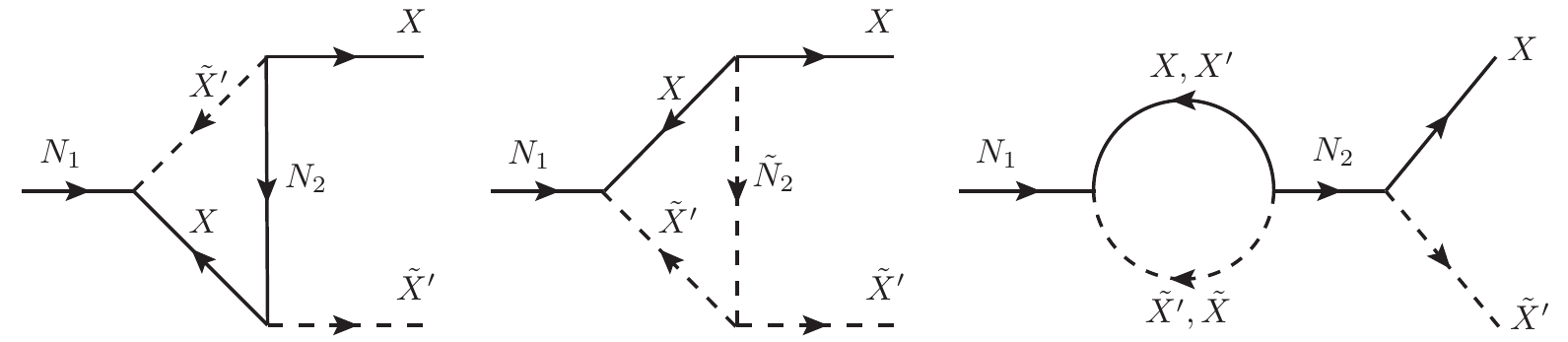}
\caption{
Loop diagrams which are responsible for the
genesis of asymmetric dark matter from the decay of $N_i$ to final states  $X\tilde X'$
and there are similar diagrams for the decay of the $N_i$ to the final states $\tilde X X'$,
and for the decay of $\tilde N_i$ to $X X'$ and to $\tilde{X} \tilde{X}'$.}
\label{genesis}
\end{center}
\end{figure}

The genesis of the dark matter asymmetry arises from the interference  of the
one-loop diagrams shown in Fig.~\ref{genesis} with the tree-level diagrams
similar to the conventional leptogenesis diagrams~\cite{Fukugita:1986hr,Murayama:1992ua,Plumacher:1996kc,Covi:1996wh}.
The asymmetries, i.e.,  the excess of
$\hat{X},\hat{X}'$ over their anti-particles $\overline{\hat{X}},\overline{\hat{X}'}$
are measured by
\begin{align}
\epsilon_{X\tilde{X}'} & =\frac{\Gamma(N_{1}\to X\tilde{X}')-\Gamma(N_{1}\to\bar{X}\tilde{X}'^*)}{\Gamma(N_{1}\to X\tilde{X}')+\Gamma(N_{1}\to\bar{X}\tilde{X}'^*)}\,,\\
\epsilon_{\tilde{X}X'} & =\frac{\Gamma(N_{1}\to\tilde{X}X')-\Gamma(N_{1}\to\tilde{X}^*\bar{X}')}{\Gamma(N_{1}\to\tilde{X}X')+\Gamma(N_{1}\to\tilde{X}^*\bar{X}')}\,,\\
\epsilon_{XX'} & =\frac{\Gamma(\tilde{N}_{1}\to XX')-\Gamma(\tilde{N}_{1}\to\bar{X}\bar{X}')}{\Gamma(\tilde{N}_{1}\to XX')+\Gamma(\tilde{N}_{1}\to\bar{X}\bar{X}')}\,,\\
\epsilon_{\tilde{X} \tilde{X}'} & =\frac{\Gamma(\tilde{N}_{1}\to \tilde{X} \tilde{X}')-\Gamma(\tilde{N}_{1}\to\tilde{X}^* \tilde{X}'^*)}{\Gamma(\tilde{N}_{1}\to \tilde{X} \tilde{X}')+\Gamma(\tilde{N}_{1}\to\tilde{X}^* \tilde{X}'^*)}\,.
\end{align}
There are two types of loops involved: vertex contribution and wave contribution as shown in Fig~\ref{genesis}.
It's straight forward to compute the asymmetry parameters $\epsilon_{X\tilde{X}'}$ etc defined above.
It turns out the contributions of the vertex diagrams and the wave diagrams satisfy the following
relations~\footnote{
The analysis assumes massless particles running in the loops aside
from the heavy field $\hat N_2$. Complete results will be presented elsewhere~\cite{wzfpn}.
}
\begin{gather}
\epsilon_{X\tilde{X}'}^{vertex}=\epsilon_{\tilde{X}X'}^{vertex}=\epsilon_{XX'}^{vertex}=\epsilon_{\tilde{X}\tilde{X}'}^{vertex}\equiv\epsilon^{vertex}\,,\\
\epsilon_{X\tilde{X}'}^{wave}=\epsilon_{\tilde{X}X'}^{wave}=\epsilon_{XX'}^{wave}=\epsilon_{\tilde{X}\tilde{X}'}^{wave}\equiv\epsilon^{wave}\,.
\end{gather}
Specifically, we have~\footnote{
The assumption of near degeneracy of the Majorana masses leads to the so-called resonant
leptogenesis~\cite{Pilaftsis:2003gt}. However, in the analysis here we do not make use of this mechanism.}
\begin{align}
\label{vw1}
\epsilon^{vertex} & =-\frac{1}{8\pi}\frac{{\rm Im}(\lambda_{1}^{2}\lambda_{2}^{*2})}{|\lambda_{1}|^{2}}\frac{M_{2}}{M_{1}}\ln\frac{M_{1}^{2}+M_{2}^{2}}{M_{2}^{2}}\,,\\
\epsilon^{wave} & =-\frac{1}{8\pi}\frac{{\rm Im}(\lambda_{1}^{2}\lambda_{2}^{*2})}{|\lambda_{1}|^{2}}\frac{M_{1}(M_{1}+M_{2})}{M_{2}^{2}-M_{1}^{2}}\,.
\label{vw2}
\end{align}
Thus the total asymmetry parameter is the sum of the vertex and the wave contributions
as given by Eqs.~\eqref{vw1}, \eqref{vw2} and one has the following equalities
\begin{equation}
\epsilon_{X\tilde{X}'}=\epsilon_{\tilde{X}X'}=\epsilon_{XX'}=\epsilon_{\tilde{X}\tilde{X}'}\equiv\epsilon\,.
\end{equation}
where $\epsilon$
is the sum of $\epsilon^{vertex}$ and $\epsilon^{wave}$ and in the
limit $M_{2}\gg M_{1}$, we obtain
\begin{equation}
\epsilon=\epsilon^{vertex}+\epsilon^{wave}\approx-\frac{1}{4\pi}\frac{{\rm Im}(\lambda_{1}^{2}\lambda_{2}^{*2})}{|\lambda_{1}|^{2}}\frac{M_{1}}{M_{2}}\,.
\label{epsi}
\end{equation}
Thus the total excess of $X,\tilde{X},X',\tilde{X'}$ over $\bar{X},\tilde{X}^{*},\bar{X}',\tilde{X'}^{*}$
generated by the decay of $\hat{N}_1$ is given by:
\beqn
\Delta n_X \equiv
(n_{\hat{X}}- n_{\overline{\hat{X}}}) + (n_{\hat{X}'} - n_{\overline{\hat{X}'}} )\,,\nonumber
\eeqn
where $\Delta n_X$ is computed to be
\beqn
\Delta n_X
= \Big[ \frac{3}{4}(\epsilon_{X\tilde{X}'}+\epsilon_{\tilde{X}X'}) +(\epsilon_{XX'}+\epsilon_{\tilde{X}\tilde{X}'})
\Big]
\frac{\kappa \zeta(3)g_N T^3 }{\pi^2}\,,\nonumber
\label{DMAsy}
\eeqn
where  $\zeta(3)\sim 1.202$. We may further write     $\Delta n_X \approx 2\kappa s \epsilon \big/ g_*$,
where $s$ is the entropy.
Here the factor of $\frac{3}{4}$ is for $N_i$ vs a factor of  1 for $\tilde N_i$,
$g_N=2$ are the degrees of freedom of the Majorana field, $g_*$ is the entropy degrees of freedom
 for MSSM where $g_* \approx 228.75$, and
$\kappa$ is a washout factor due to inverse processes
$X+\tilde X', \tilde X+ X' \to N$ and
$X+ X', \tilde X+ \tilde X' \to \tilde N$.
A computation of $\kappa$
requires solving the Boltzman equations~\cite{Buchmuller:2002rq}.
In our analysis here we set $\kappa =0.1$.
The excess of $\hat{X}, \hat{X}'$ then give rise to a non-vanishing $(B-L)$-number in the early universe:
\begin{equation}
(B-L)_t = (+1)\times \Delta n_X
= 2\kappa s \epsilon \big/ g_*\,,
\label{epsB-L}
\end{equation}
where $(B-L)_t$ is the total $B-L$ in the Universe. \\

{\it Leptogenesis and Baryogenesis from dark matter:}
We consider now supersymmetric interactions which can transfer a $B-L$ asymmetry
from the dark sector to the visible sector, which give rise to both
leptogenesis and baryogenesis.
Leptogenesis is accomplished via
a transfer equation
similar to the one  adopted in previous works~\cite{Kaplan:2009ag}
\beqn
W_{\rm tran} = \frac{1}{M_{\rm asy}^n} \hat{X}\hat{X}' {\cal O}^{\rm MSSM}_{\rm asy}\,.
\eeqn
As a specific example we consider the case
\beqn
W_{\rm tran} = \frac{1}{M_{\rm asy}^2} \hat{X}\hat{X}' (L H_u)^2\,.
\label{asy}
\eeqn
We assume that
the transfer interaction is active only above the temperature $T_{\rm int}$
which lies above $T_{{\rm SUSY}}$
which corresponds to the SUSY breaking scale,
i.e., $T_{{\rm int}}>T_{{\rm SUSY}}\sim~{\cal O}({1~\rm TeV})$.
Of course there are other possible choices for $T_{\rm int}$ where
$W_{{\rm tran}}$ would decouple (see~\cite{Feng:2012jn} for a comprehensive
discussion).
To determine the generation of lepton number and baryon number from the couplings of the dark sector to
the visible sector we use the standard thermal equilibrium method introduced in~\cite{Harvey:1990qw}.
At the scale $T_{\rm int}> T_{\rm SUSY}$ all the MSSM fields are ultra-relativistic and
have non-vanishing chemical
potential while the gauge bosons have a vanishing chemical potential.
Constraints on the chemical potential arise from a variety of sources.
The superpotential in MSSM reads
\begin{equation}
W_{\rm MSSM}=g_{u}QH_{u}U^{c}-g_{d}QH_{d}D^{c}-g_{e}LH_{d}E^{c} + \mu H_uH_d
\end{equation}
where the Yukawa sector gives the following constraints on the chemical potentials
\begin{gather}
\mu_{H_{d}}=-\mu_{L}-\mu_{E^{c}}=-\mu_{Q}-\mu_{D^{c}}\,,\\
\mu_{H_{u}}=-\mu_{Q}-\mu_{U^{c}}\,,
\end{gather}
while the Higgs mixing term $\mu H_{u}H_{d}$ gives
\begin{equation}
\mu_{H_{u}}+\mu_{H_{d}}=0\,.
\end{equation}
Additionally, the sphaleron processes ($\prod Q_{i}Q_{i}Q_{i}L_{i}$, $i=1,2,3$)
give us the constraint,
\begin{equation}
3\mu_{Q}+\mu_{L}=0\,,\label{SPHEQ-1}
\end{equation}
and the  condition that the total hypercharge of the Universe
is zero leads to%
\begin{equation}
Y =
 3\times(3\mu_{Q}-6\mu_{U^{c}}+3\mu_{D^{c}}-3\mu_{L}+3\mu_{E^{c}}+2\mu_{H_{u}})=0\,.
\end{equation}
Solving above equations, we can express all the chemical potentials
in terms of the chemical potential of one single field, i.e., $\mu_{L}$.
The solutions show a generation of lepton and baryon number in the visible sector and one has
\begin{align}
\label{B}
&(B-L)_v  =-\tfrac{237}{7}\mu_L\,,
\end{align}
where $(B-L)_v$ is the $B-L$ in the visible sector.\\

The dark matter also undergoes a readjustment as a consequence of thermal equilibrium
and the residual  dark matter after the action of the asymmetry transfer interaction Eq.~\eqref{asy} can be gotten via the chemical potential equation:
\begin{equation}
\mu_{\hat X} +\mu_{\hat{X}'}+2\,(\mu_{L}+\mu_{H_{u}})  =0\,.
\end{equation}
Also the mass terms of $\hat X, \hat{X}^c, \hat X', \hat{X}'^c$ in Eq.~\eqref{Wgen} at equilibrium give
\begin{equation}
\mu_{\hat{X}} + \mu_{\hat{X}^c} = \mu_{\hat{X}'} +\mu_{\hat{X}'^c} = 0 \,,
\end{equation}
which lead to
\begin{equation}
\mu_{\hat{X}}+\mu_{\hat{X}'}= -\mu_{\hat{X}^c} -\mu_{\hat{X}'^c} = -\frac{22}{7}\mu_{L}\,.
\end{equation}
From the above we find
\beqn
\sum_i X_i = -\frac{3\times 2\times 22}{7}\mu_L = \frac{44}{79} (B-L)_v\,.
\label{X/B-L}
\eeqn
where $X_i$ are the number of dark matter particles of species $i$ (in this case $i$ takes on values
from $1-4$ since we have 4 species: $\hat X, \hat{X}^c, \hat X', \hat{X}'^c$).
In Eq.~(\ref{X/B-L}) the factor of $3 = 1 + 2$ counts the chemical potentials
for both fermions and bosons of the superfields
(from converting the $\mu$'s to the excess of the number densities),
and the factor of 2 counts the contribution also from $\hat{X}^c, \hat{X}'^c$.
Using Eq.~\eqref{X/B-L} it is also easy to find the ratio that
\begin{equation}
\frac{(B-L)_v}{(B-L)_t} \approx 0.64\,.
\label{v/t}
\end{equation}

We can now determine the dark matter mass
from the ratio of the dark matter relic density to baryonic relic density. Thus the
 ratio of dark matter relic density to the baryonic matter density is given by
\beqn
\frac{\Omega_{\rm DM}}{\Omega_{\rm B}} = \frac{\sum_{i} X_i \cdot m_{\rm DM}^i}{B \cdot m_{\rm B}}\approx 5\,,
\label{DMMass}
\eeqn
where  $m_{\rm DM}^i$ are the masses of the dark matter particles and
$m_{\rm B}\sim 1$ GeV.
There is an important subtlety here that although the total dark particle number is fixed after the
asymmetry transfer interaction decouples, the total baryon number, however, keeps changing
because of the sphaleron processes. As was explained in~\cite{Feng:2012jn},
the total baryon number to be used in this formula is $B_{\rm final}$ after
the sphaleron processes decouple.
Using Eq.~\eqref{DMMass} we have
\beqn
m_{\rm DM} = 5 \cdot \frac{B_{\rm final}}{\sum_i X_i}\,,
\eeqn
where
\begin{equation}
B_{\rm final} = \frac{30}{97} (B-L)_v \approx 0.31 (B-L)_v\,.
\label{BF/B-L}
\end{equation}
Thus we obtain
\beqn
m_{\rm DM} \approx 2.78~{\rm GeV}\,.
\eeqn
From astrophysical constraints one has~\cite{Beringer:1900zz}
\begin{equation}
B_{\rm final} / s \sim  6  \times 10^{-10}\,.
\end{equation}
Using Eqs.~\eqref{epsB-L}, \eqref{v/t} and \eqref{BF/B-L}, we obtain
\begin{equation}
\epsilon \sim 4\times 10^{-6}\,,
\label{estimate}
\end{equation}
which sets bounds for complex couplings $\lambda_i$ and the ratio $M_1/M_2$.
This can be seen by noting that Eq.~\eqref{epsi} can be written in the form
\begin{equation}
\epsilon \approx-\frac{M_1}{4\pi M_2} |\lambda_{1}|^{2} \sin{2\alpha}\,.
\end{equation}
where $\alpha$ is the relative phase of $\lambda_1$ and $\lambda_2$. Thus very reasonable choices
of the parameters, such as $M_1/M_2 \sim |\lambda_1|\sim\alpha \sim 10^{-1}$,
lead to consistency with Eq.~\eqref{estimate}.
We note that since we are considering a $U(1)_x$ gauge symmetry,
the Majorana mass terms for the dark particles are forbidden.
Thus, the dark matter asymmetry generated in the early universe would not be washed out
by oscillations.
 The dissipation of the symmetric component of
dark matter can be achieved by gauge kinetic energy mixing~\cite{Holdom:1985ag}
of $U(1)_x$ and $U(1)_Y$
and via \st mass mixing~\cite{Kors:2004dx,Cheung:2007ut,Feldman:2007wj}
Thus dissipation of the thermally produced
 $X, X'. \tilde X, \tilde X'$  and their anti-particles
occurs  from their annihilation via the
$Z'_x$ boson exchange coupled with a Breit-Wigner pole
enhancement~\cite{Griest:1990kh,Gondolo:1990dk,Arnowitt:1993mg,Feldman:2008xs,Feng:2012jn}.
We are able to deplete sufficient amounts of the symmetric  component of dark matter
(so it is less than $10\%$ of the total dark matter relic density)
with a mixing between $U(1)_x$ and $U(1)_Y$ as low as $\delta \sim 0.001$~\cite{Feng:2012jn}
(where $\delta$ is the mixing angle) in the desired mass region of dark matter.
We  also note that the $U(1)_x$ gaugino $\lambda_x$  is given
a soft mass $\mathcal{L}_{\lambda_x} = m_{\lambda} \bar \lambda_x \lambda_x$.
It can then decay into $X\tilde X$ or $X' \tilde X'$ via the supersymmetric interaction
$\mathcal{L} \sim \lambda_x X \tilde X+ \lambda_{x} X' \tilde X' + h.c.$,
where we  assume  $m_{\lambda} > m_X + m_{\tilde X}$.
Thus the gaugino $\lambda_x$ decays into dark particles and is removed from the
low energy  spectrum.\\


{\it Phenomenology}: We give now further discussion of the phenomenological aspects of the model
with more details.
First we note that the Lagrangian with kinetic mixing between two gauge fields $A_{1\mu}, A_{2\mu}$
corresponding to the gauge  groups $U(1)_x$ and $U(1)_Y$ where the mass of one of the field arises
from the \st mechanism is given by
$\mathcal{L} =\mathcal{L}_0 + \mathcal{L}_m + \mathcal{L}_1 $ where
\beqn
\mathcal{L}_0 &=&
    - \tfrac{1}{4}F_{1\mu\nu}F_1^{\mu\nu}
    - \tfrac{1}{4}F_{2\mu\nu}F_2^{\mu\nu}
    - \tfrac{\delta}{2}F_{1\mu\nu}F_2^{\mu\nu}\,,\nonumber\\
\mathcal{L}_m &=& -\tfrac{1}{2} M^2 A_{1\mu} A^{\mu}_1\,,\nonumber\\
\mathcal{L}_1 &=&
     J'_{\mu}A_1^{\mu}
    +J_{\mu}A_2^{\mu}\,.
       \label{onlykt}
\eeqn
We make a transformation to bring  kinetic energy term in its canonical form
using the transformation
\begin{eqnarray}
  \left[\begin{array}{c}
    A_1^{\mu}\\
    A_2^{\mu}
         \end{array} \right]
         \to  K_0
            \left[\begin{array}{c}
    Z^{\mu '}\\
    B^{\mu }
         \end{array} \right] \,.
\end{eqnarray}
where $K_0$ has the form
\begin{eqnarray}
 K_0=  \left[\begin{array}{cc}
  \frac{1}{\sqrt{1-\delta^2}} & 0 \\
  \frac{-\delta}{\sqrt{1-\delta^2}}    & 1\\
         \end{array} \right]\,.
\end{eqnarray}
The interaction Lagrangian in the new basis is given by
\begin{eqnarray}
{\cal{L}}_{1} &= & \left( \frac{-\delta}{\sqrt{1-\delta^2}} J_{\mu} +
       \frac{1}{\sqrt{1-\delta^2}} J_{\mu}' \right) Z'^{\mu}
            +              J_{\mu} B^{\mu}\,.
\end{eqnarray}
We identify $J_{\mu}$ with the hypercharge current and $J_{\mu}'$ with the
current arising from the dark sector to  which $Z'$ couples.
\\

One of the important phenomenological consequences of the above is that
the photon does not couple to the dark sector and thus the dark matter carries no milli-charge
which is in contrast to models where the mixing between the two $U(1)$'s, one in the visible and
the other in the dark sector,
occurs via the \st mechanism. Consequently there are no experimental
constraints on the mixing parameter $\delta$ arising from the experimental limits
on milli-charges.\\

One of the strongest experimental constraints on the $Z'$ mass  and its coupling to the visible sector comes from corrections to
$g_{\mu}-2$. The current experimental limit on the deviation from the Standard Model result is
given by the Brookhaven experiment so that~\cite{Beringer:1900zz}
\beqn
\Delta (\frac{g_{\mu}-2}{2}) < 3 \times 10^{-9}\,.
\eeqn
Now in the current model the correction to $g_\mu-2$ at the one loop order is given by
\beqn
\Delta (g_\mu-2) = \frac{\delta^2}{1-\delta^2} \frac{g_Y^2 C m_\mu^2} { 24 \pi^2   M_{Z'}^2}\,,
\eeqn
where $C=2 Y_L Y_R$, with $Y_L=-1/2$ and $Y_R=-1$ ($Y'$'s are normalized so that
$T_3+Y= Q$).
Using the input
$\delta \simeq  0.001$ and $M_{Z'}\simeq 10$~GeV  one finds that
$\Delta (g_{\mu}-2) \sim 10^{-14}$ and thus the $Z'$ exchange makes a negligible
contribution to $\Delta(g_{\mu}-2)$.
Further the LEP II constraints on the $Z'$ couplings imply that~\cite{LEP:2003aa}
\beqn
M_{Z'}/g_{Z'ff} > 6 ~{\rm TeV}\,,
\label{lep2}
\eeqn
where $g_{Z'ff}\equiv g_Y \sqrt C (\delta/\sqrt{1-\delta^2})$. In deducing the $6$~TeV limit in Eq.~\eqref{lep2}
we have used
the $\Lambda_{VV}^+$ value of $21.7$~TeV in~\cite{LEP:2003aa}.
Using the same inputs as above gives
for the left hand side  of Eq.~\eqref{lep2} the result $\simeq 28$~TeV which adequately satisfies Eq.~\eqref{lep2}.
\\

One important aspect of this model relative to other models, is that it presents a multi-component picture
of dark matter. Thus as mentioned above the dark matter consists of the leptonically charged matter
consisting of $X,X', X^c, X'^{c}$ as well as the conventional supersymmetric LSP with R-parity, i.e., the
neutralino. For the cosmic coincidence picture to work (i.e., the ratio of dark matter to baryonic matter to
be $\sim 5$) the symmetric component of leptonic dark matter must be depleted so that it is no more than
a small fraction of the total leptonic dark matter, i.e., that the leptonic dark matter is mostly the asymmetric
dark matter. At the same time the relic density of the LSP neutralino should also not exceed a small
fraction of the total relic density of dark matter. It is possible to achieve both these features in this model.
The analysis of this part is similar to the analysis given in~\cite{Feng:2012jn}.
The total relic density consists of
\beqn
\Omega_{\rm DM} = \Omega_{\psi} + \Omega_{\bar \psi} + \Omega_{\tilde \chi^0}\,,
\eeqn
where $\Omega_{\psi}= m_{\psi} n_{\psi}/\rho_c$ and $\rho_c$ is the critical matter density of the Universe,
and similar relation holds for $\Omega_{\bar \psi}$ with ${\psi}$ replaced with ${\bar \psi}$.
The analysis of the relic densities $\Omega_\psi$ and $\Omega_{\bar \psi}$ is very similar to the
one given in~\cite{Feng:2012jn} and one finds that the symmetric component can be depleted to less than
10\% of the asymmetric part.
An analysis of the relic density from asymmetric dark matter requires
solution to the Boltzmann equations which contain the asymmetry.
The presence of the asymmetry further helps to deplete the symmetric component of the dark matter.
A more in depth discussion of this topic can be found in~\cite{Feng:2012jn}.
Further, there exists a significant part of the parameter space of
MSSM where the  relic density of neutralinos can be  10\% or less of the current relic density.
The analysis of~\cite{Feng:2012jn} shows that even with 10\% of the relic density the neutralino dark
matter would be accessible in dark matter searches.
The above also offers a direct test of the model in dark matter searches. For example,
suppose we observe dark matter in direct dark searches which can be fitted within MSSM with
a certain set of assumed soft parameters which, however, give only one tenth of the relic density.
This is precisely the phenomenon that the current model can explain.
However, the leptonic dark matter would be difficult
to see in direct searches for dark matter as well as in collider experiments because of its small couplings
to the visible sector via the $Z'$ exchange.
However, future colliders with higher sensitivity and accuracy could have
the possibility to explore the $Z'$ gauge boson with tiny couplings to the Standard Model particles.
\\

We discuss now the differences between our work and the previous works~\cite{haba,falk,chun}.
The major difference is that in our work the primordial fields in the early universe
decay only into the dark sector, while in the previous works~\cite{haba,falk,chun}
the heavy fields (either inflaton or right-handed neutrino) decay to both visible and dark sector particles.
Thus in our work the asymmetry at the beginning was created only
in the dark sector and then transmuted into the visible matter,
while in the previous works~\cite{haba,falk,chun} the asymmetry is generated simultaneously
in both the dark sector and the visible sector.
In addition, in our model the dark particles carry $U(1)_x$ gauge charges,
which forbids the dangerous Majorana mass terms that would generate oscillations
of the dark particles and their anti-particles which could washout the asymmetry~\cite{Buckley:2011ye}.
This is a feature which is not necessarily shared by all the models of~\cite{haba,falk,chun}
(see \cite{haba}).  In our model the $U(1)_x$ mixes kinetically with the hypercharge and
the symmetric component of dark matter annihilates through a $Z'$ pole into the Standard Model particles.
This mechanism of annihilation of the symmetric component of dark matter
is very different with the one in previous works~\cite{haba,falk,chun}.
Another feature that differentiates our work with those of~\cite{haba,falk,chun} is that in our model the dark particles are predicted to be around 3~GeV,
while in the  works of~\cite{haba,falk,chun} the mass of dark particles can vary in a wide range.
Finally, another distinguishing feature of our model is the multicomponent nature of dark matter consisting of dark
particles carrying leptonic charges as well as a small fraction of neutralinos  which could still be
detectable in dark matter searches.
\\


{\it Conclusion:\label{d}}
In this  work
we have discussed the possibility that the decay of the primordial fields create asymmetric dark matter,
and the lepton and baryon excess arise as a consequence of transmutation of the asymmetric dark matter.
The symmetric component of dark matter is depleted via
kinetic mixing between $U(1)_x$ and the hypercharge gauge group and hence annihilates to the Standard Model particles.
The Majorana mass terms for  the dark particles are forbidden
since they carry $U(1)_x$ gauge charges
hence the asymmetric dark matter generated in the early universe would not be washed out
by oscillations and thus sources leptogenesis and baryogenesis.
The model accomplishes three things: it provides a framework for (i) baryogenesis, (ii) generation of dark matter,
and (iii) an explanation of cosmic coincidence, i.e., $\Omega_{\rm DM}/\Omega_{\rm B}\sim 5$.
The model, however, allows for a small fraction of dark matter ($\sim 10\%$)  to be  neutralinos
which nevertheless can be detected in direct searches such as in XENON-1T experiment~\cite{Aprile:2012zx}.
The model contains a new $Z'$ gauge boson which couples to both the dark particles and Standard Model particles,
with mass around $10~{\rm GeV}$.
Its mass and couplings are consistent with the Brookhaven $g_\mu-2$ experiment and with the LEP constraints.
However, more sensitive future colliders with sensitivity
better than a factor of about 10 should be able to detect this vector boson
and test its couplings.
\\

{\it Acknowledgments:}
The work of PN is supported in part by the U.S. National Science Foundation (NSF) grants
PHY-0757959 and  PHY-070467. WZF is supported by funds from The Hong Kong University of Science and Technology.
AM is supported by the STFC grant ST/J000418/1. WZF and AM would like to thank Northeastern University for hospitality,
and WZF is thankful to HaiPeng An  for helpful discussions.

\end{document}